\documentclass[11pt]{article}
\usepackage{amssymb,latexsym,amsmath}
\usepackage{amsfonts, graphics}
\newtheorem{theorem}{Theorem}

\begin{document}
\title{On the Buchdahl inequality for spherically symmetric static shells}
\author{H{\aa}kan Andr\'{e}asson\\
        Department of Mathematics, Chalmers,\\
        S-41296 G\"oteborg, Sweden\\
        email\textup{: \texttt{hand@math.chalmers.se}}}

\maketitle

\begin{abstract}
A classical result by Buchdahl \cite{Bu1} shows that for static
solutions of the spherically symmetric Einstein-matter system, the total ADM
mass M and the area radius R of the boundary of the body, obey the
inequality $2M/R\leq 8/9.$ The proof of this inequality rests on the hypotheses
that the energy density is non-increasing outwards and that the pressure is
isotropic. In this work neither of Buchdahl's hypotheses are assumed. 
We consider non-isotropic spherically symmetric shells, supported 
in $[R_0,R_1], R_0>0,$ 
of matter models for which the energy density $\rho\geq 0,$ and 
the radial- and tangential pressures $p\geq 0$ and $q,$ satisfy 
$p+q\leq\Omega\rho, \;\Omega\geq 1.$ We show a Buchdahl type inequality 
for shells which are thin; given an $\epsilon<1/4$ there is a $\kappa>0$ 
such that $2M/R_1\leq 1-\kappa$ when $R_1/R_0\leq 1+\epsilon.$ 
It is also shown that for a sequence of solutions such that $R_1/R_0\to 1,$ 
the limit supremum of $2M/R_1$ of the sequence is bounded by 
$((2\Omega+1)^2-1)/(2\Omega+1)^2.$ In particular if $\Omega=1,$ which 
is the case for Vlasov matter, the boumd is $8/9.$ 
The latter 
result is motivated by numerical simulations \cite{AR2} which indicate that 
for non-isotropic shells of Vlasov matter $2M/R_1\leq 8/9,$ and moreover, that 
the value $8/9$ is approached for shells with $R_1/R_0\to 1$. 
In \cite{An2} a sequence 
of shells of Vlasov matter is constructed with the properties that $R_1/R_0\to 1,$ and that 
$2M/R_1$ equals $8/9$ in the limit. We emphasize that in the present paper no field 
equations for the matter are used, whereas in \cite{An2} the Vlasov equation is important. 
\end{abstract}
\section{Introduction}
The metric of a static spherically symmetric spacetime takes the following form in Schwarzschild coordinates 
\begin{displaymath}
ds^{2}=-e^{2\mu(r)}dt^{2}+e^{2\lambda(r)}dr^{2}
+r^{2}(d\theta^{2}+\sin^{2}{\theta}d\varphi^{2}),
\end{displaymath}
where $r\geq 0,\,\theta\in [0,\pi],\,\varphi\in [0,2\pi].$
Asymptotic flatness is expressed by the boundary conditions
\begin{displaymath}
\lim_{r\rightarrow\infty}\lambda(r)=\lim_{r\rightarrow\infty}\mu(r)
=0,
\end{displaymath}
and a regular centre requires
\begin{displaymath}
\lambda(0)=0.
\end{displaymath}
The Einstein equations read 
\begin{eqnarray}
&\displaystyle e^{-2\lambda}(2r\lambda_{r}-1)+1=8\pi r^2\rho,&\label{ee12}\\
&\displaystyle e^{-2\lambda}(2r\mu_{r}+1)-1=8\pi r^2
p,&\label{ee22}\\
&\displaystyle
\mu_{rr}+(\mu_{r}-\lambda_{r})(\mu_{r}+\frac{1}{r})= 4\pi
qe^{2\lambda}.&\label{ee4}
\end{eqnarray}
Here $\rho$ is the energy density, $p$ the radial pressure and $q$ is the 
tangential pressure. If the pressure is isotropic, i.e. $2p=q,$ a solution
will satisfy the well-known Tolman-Oppenheimer-Volkov equation 
for equilibrium 
\begin{equation}
p_{r}= -\mu_{r}(p + \rho).\label{TOV1} 
\end{equation}
In the case of non-isotropic pressure this equation generalizes to 
\begin{equation}
p_{r}= -\mu_{r}(p + \rho) - \frac{1}{r}(2p-q).\label{TOV2}
\end{equation}
In 1959 Buchdahl \cite{Bu1} showed that static fluid spheres satisfy the inequality $$2M/R\leq \frac{8}{9},$$ where $M$ is the ADM mass,
$$M=\int_0^{\infty}4\pi\eta^2 \rho(\eta)d\eta,$$ 
and $R$ is the outer boundary (in Schwarzschild coordinates) of the fluid sphere. The fluid spheres considered by Buchdahl have isotropic pressure and are in addition assumed to have an energy density which is non-increasing outwards. Note that the isotropy assumption implies that also the pressure is monotonic. This follows from (\ref{TOV1}) since $\mu_r\geq 0,$ which is a consequence of (\ref{ee12}) and (\ref{ee22}). It is sometimes argued that the assumption of non-increasing energy density is natural in the sense that the fluid sphere is unstable otherwise \cite{W}. However, at least for Vlasov matter this is certainly not the case. The existence of stable, spherically symmetric static shells of Vlasov matter (i.e. the matter is supported in $[R_0,R_1],\;R_0>0$) has been demonstrated numerically eg. in \cite{AR1}. Moreover, these shells are in general non-isotropic. Static solutions of Vlasov matter can also be constructed which do not have the shell structure \cite{RR4}, i.e. $R_0=0.$ Also for these solutions, the hypotheses of non-increasing energy density and isotropic pressure are in general violated. The method of proof by Buchdahl rests on the monotonicity of both $\rho$ and $p,$ and a natural question is then if there is a Buchdahl type inequality also for static solutions where neither $\rho$ nor $p$ is monotonic? 
Quite surprisingly, numerical results in \cite{AR2} support that $2M/R_1<8/9$ for any static solution of the Einstein-Vlasov system and that there are solutions with $2M/R_1$ arbitrary close to $8/9$. ($R_1$ will always denote the outer boundary of the static solution.) An analytic investigation for Vlasov matter is given in \cite{An2}. 
In this work we will however consider static shells of any matter model (which has static solutions) for which $\rho$ and $p$ are non-negative and 
\begin{equation}
p+q\leq\Omega\rho, \;\Omega\geq 1.\label{Omega}
\end{equation}
In particular, by taking $\Omega=3,$ any matter model with $\rho$ and $p$ non-negative, which satisfies the dominant energy condition is admitted. Note that the condition that $\rho\geq 0$ can be replaced by requiring that the weak energy condition is satisfied. In the case of Vlasov matter $\rho, p$ and $q$ are all non-negative and (\ref{Omega}), with strict inequality, is satisfied with $\Omega=1.$ \\

\textit{Remark: }The condition that $p\geq 0$ can be replaced by $-p\leq c\rho, \;0<c<1,$ and $\Omega>0$ is sufficient in order to derive a Buchdahl type inequality, but the formulation of the results and their proofs would be slightly more cumbersome so we have chosen to use the conditions above. \\

Bondi \cite{Bo} has investigated (not rigorously) if isotropic solutions, without the assumption of non-increasing energy density, obey a Buchdahl type inequality. He considers models for which $\rho\geq 0,\; \rho\geq p,$ or $\rho\geq 3p,$ and gets $0.97,\; 0.86$ and $0.70$ respectively as upper bounds of $2M/R_1.$ The isotropic condition is crucial though, since, as a consequence of \cite{An2} (and numerically in \cite{AR2}), the second and third bounds are violated by non-isotropic steady states of Vlasov matter. Note that $\rho\geq p$ always holds for Vlasov matter. 

\textit{Remark: }We point out that steady states of Vlasov matter which are isotropic and which have non-increasing energy density do exist but the analysis of such states is identical to Buchdahl's original analysis.\\ 

Non-isotropic solutions have previously been studied in \cite{BR} where it is shown that $2M/R_1<1,$ and in \cite{F} static shells where the density is concentrated at a single radius and the source in the Einstein equations is distributional, were found to satisfy $2M/R_1\leq 24/25.$  
Here we will consider shells which are thin in the sense that $R_1/R_0\leq 1+\epsilon,$ and we show that for any $\epsilon<1/4$ there is a $\kappa>0$ such that $2M/R_1\leq 1-\kappa.$ We also show that for a sequence of static shells such that $R_1/R_0\to 1,$ the limit supremum of $2M/R_1$ is bounded by $8/9.$ 

The outline of the paper is as follows. 
In the next section our main results are presented in detail and the proofs are given in section 3. 
\section{Main results}
\begin{theorem}
Let $a>0,\; K\geq 5/4+3\Omega/4,$ and let $$\kappa=\frac{1}{(2\Omega+1+\frac{2(\Omega+1)^2}{2\Omega+1}(K+a))^2}.$$ Consider a solution to the Einstein equations such that $p+q\leq\Omega\rho,$ with support in $[R_0,R_1],$ and such that $R_1/R_0\leq (1+\epsilon),$ where $$\epsilon < \min{\{\frac{1-(2\Omega+3)\kappa-(\Omega+1)(1-\kappa)e^{-4K/9}}{2(\Omega+1)},a\sqrt{\kappa}\}}.$$ 
Then 
\[
2M/R_1\leq 1-\kappa, 
\]
and 
\[ 
\sup_{r}\frac{2m}{r}\leq 1-\kappa/2, \mbox{  if  }\epsilon\leq\kappa/2.
\]
\end{theorem} 
The condition on $K$ is to ensure that the first term in the expression for $\epsilon$ is positive and can be sharpened. 
The numbers $\kappa$ and $\epsilon$ are related in the sense that a larger $\epsilon$ can be chosen if $\kappa$ is taken smaller and vice versa. 
In view of the next theorem, which gives a much improved value of $\kappa$ when $\epsilon$ is made arbitrary small, it seems more interesting to make $\epsilon$ large even if $\kappa$ then becomes smaller, since it gives us an estimate on the possible thickness of the shells that our method allows in order to admit a Buchdahl inequality at all. We see immediately from the formula above that our method cannot handle $\epsilon\geq 1/4,$ but by fixing $K$ large and then taking $a$ sufficiently large we can have $\epsilon$ arbitrary close to $1/4.$ A choice which favours a large value of $\epsilon$ rather than a large $\kappa,$ in the case when $\Omega=1,$ is $K=6$ and $a=9,$ which implies that when $\epsilon<1/5$ it holds that $2M/R_1\leq 1-1/43^2.$ 

Next we consider a sequence of shells supported in $[R_0^j,R_1^j],$ such that $R_1^j/R_0^j\to 1.$ We find that, as the support gets thinner, the value $$\frac{(2\Omega+1)^2-1}{(2\Omega+1)^2}$$ of $2M/R_1$ can not be exceeded. In particular, if $\Omega=1,$ then the same bound $8/9$ as in Buchdahl's original work is obtained, and if $\Omega=2,$ then the bound is $24/25$ which agrees with the value found in \cite{F}. The latter agreement is not surprising since the case considered in \cite{F} is an infinitely thin shell with radial pressure zero which satisfies the dominant energy condition. In our terminology this means $\Omega=2.$ The former result is more surprising since the steady state that realizes $8/9$ in Buchdahl's original work is the well-known interior solution with constant energy density which is completely different from our situation where the energy density gets more and more peaked at the radius $R_1.$ 
\begin{theorem}
Assume that $\{(\rho^{j},p^j,q^j,\mu^j)\}_{j=1}^{\infty}$ is a sequence of solutions to the static Einstein-matter system such that $p^j+q^j\leq\Omega\rho^j,$ with support in $[R_0^j,R_1^j],$ and such that 
\begin{equation}
\lim_{j\to\infty}\frac{R_1^j}{R_0^j}=1.\label{hypothesis}
\end{equation}
Then 
\begin{equation}
\limsup_{j\to\infty}\frac{2M^j}{R_1^j}\leq\frac{(2\Omega+1)^2-1}{(2\Omega+1)^2}.\label{limit} 
\end{equation}
\end{theorem}
The hypothesis (\ref{hypothesis}) holds for a class of static shell solutions of the Einstein-Vlasov system \cite{An2}. Numerical simulations \cite{AR2} (where $\Omega=1$) also indicate that for a sequence of shell solutions for which $2M/R_1$ increase and take as large values as possible ($<8/9$) the hypothesis (\ref{hypothesis}) is satisfied. In \cite{An2} a sequence of shells of Vlasov matter with the property (\ref{hypothesis}) is indeed constructed which \textit{attains} $8/9$ in the limit and thus improves statement (\ref{limit}) for Vlasov matter. 
\section{Proofs of Theorem 1 and 2}
Before starting to prove the theorems let us collect a couple of facts concerning the system (\ref{ee12})-(\ref{ee4}). A consequence of equation (\ref{ee12}) is that 
$$e^{-2\lambda}=1-\frac{2m(r)}{r},$$ and from (\ref{ee22}) it then follows that 
$$\mu_r=(\frac{m}{r^2}+4\pi rp)e^{2\lambda}.$$ 
Adding (\ref{ee12}) and (\ref{ee22}) and using the boundary conditions at $r=\infty$ gives
\begin{equation}
\mu(r)+\lambda(r)=-\int_r^{\infty}4\pi\eta(\rho+p))e^{2\lambda}d\eta.
\end{equation}
In particular if $R_1$ is the outer radius of support of the matter then $$e^{\mu(r)+\lambda(r)}=1,$$ when $r\geq R_1.$ Hence, 
$$e^{\mu(r)}=e^{-\lambda(r)}=\sqrt{1-2m(r)/r},\; r\geq R_1.$$ 
The facts above will be used in the proofs without further comments. 
Next we note that the generalized Tolman-Oppenheimer-Volkov equation (\ref{TOV2}) implies that a solution satisfies 
\begin{equation}
(\frac{m}{r^2}+4\pi rp)e^{\mu+\lambda}=\frac{1}{r^2}\int_0^r 4\pi\eta^2 e^{\mu+\lambda}(\rho+p+q)d\eta.\label{fundamentaleq}
\end{equation}
Indeed, let $\psi=(m+4\pi r^3p)e^{\mu+\lambda}.$ Using (\ref{TOV2}) we get 
$$\psi_r=4\pi r^2(\rho+p+q)e^{\mu+\lambda},$$ and the claim follows since $\psi(0)=0.$ \\ 
We have chosen to treat Theorem 2 before Theorem 1. \\
\textit{Proof of Theorem 2: }
First we will show that a uniform bound on $$e^{\lambda}=\frac{1}{\sqrt{1-2m/r}},$$ will imply (\ref{limit}). Hence, given $1>\kappa >0,$ assume that 
\begin{equation}
\sup_{\eta}\frac{2m^{j}(\eta)}{\eta}\leq 1-\kappa, \mbox{ for all }j.\label{uniform}
\end{equation}
Given $\epsilon >0$ we choose $j$ sufficiently large so that $$\frac{R_1^j}{R_0^j}-1\leq \epsilon.$$ 
Below we drop the index $j.$ 
Since $p(R_1)=0,\; m(R_1)=M$ and $e^{(\mu+\lambda)(R_1)}=1,$ we get by taking $r=R_1$ in (\ref{fundamentaleq}) and using that $f$ is supported in $[R_0,R_1],$ 
\begin{equation}
M=\int_{R_0}^{R_1} 4\pi\eta^2 e^{\mu+\lambda}(\rho+p+q)\,d\eta \leq (\Omega+1)\int_{R_0}^{R_1}4\pi\eta^2 e^{\mu+\lambda}\rho\, d\eta.\label{Minequality} 
\end{equation}
Here we also used that $p+q\leq\Omega\rho.$ Now $\mu$ is increasing so the right hand side is less or equal to 
\begin{eqnarray*}
&\displaystyle (\Omega+1)e^{\mu(R_1)}R_1\int_{R_0}^{R_1}4\pi\eta e^{\lambda}\rho d\eta &\\
&\displaystyle =(\Omega+1)e^{\mu(R_1)}R_1 \int_{R_0}^{R_1}(4\pi\eta\rho(\eta)-\frac{m(\eta)}{\eta^2})e^{\lambda} d\eta&\\
&\displaystyle +(\Omega+1)e^{\mu(R_1)}R_1\int_{R_0}^{R_1}\frac{m(\eta)}{\eta^2}e^{\lambda} d\eta =:S_1 +S_2.& 
\end{eqnarray*}
Since $$\lambda_r=(4\pi r\rho(r)-\frac{m(r)}{r^2})e^{2\lambda},$$ we get 
\[
S_1=-(\Omega+1)e^{\mu(R_1)}R_1 \int_{R_0}^{R_1}\frac{d}{d\eta}e^{-\lambda}\, d\eta=
(\Omega+1)e^{\mu(R_1)}R_1(1-\sqrt{1-2M/R_1}).
\]
Using that $e^{\mu(R_1)}=\sqrt{1-2M/R_1},$ we thus get 
\[
S_1=(\Omega+1)\sqrt{1-2M/R_1}R_1(1-\sqrt{1-2M/R_1})=\frac{2(\Omega+1)M\sqrt{1-2M/R_1}}{1+\sqrt{1-2M/R_1}}.
\]
The idea is now to show that $S_2$ is approaching zero as $j$ tends to infinity, since note that if $S_2=0$ we have 
\begin{equation}
M\leq\frac{2(\Omega+1)M\sqrt{1-2M/R_1}}{1+\sqrt{1-2M/R_1}},\label{original} 
\end{equation}
or 
\[
1\leq (2\Omega+1)\sqrt{1-2M/R_1},
\]
which implies that 
\[
\frac{2M}{R_1}\leq\frac{(2\Omega+1)^2-1}{(2\Omega+1)^2}. 
\]
To estimate $S_2$ we use (\ref{uniform}) and that $e^{\lambda}=(1-2m/r)^{-1/2}$ to obtain 
\begin{eqnarray}
&\displaystyle S_2 =(\Omega+1)e^{\mu(R_1)}R_1\int_{R_0}^{R_1}\frac{m(\eta)}{\eta^2 \sqrt{1-2m(\eta)/\eta}} d\eta&\nonumber\\
&\displaystyle \leq (\Omega+1)e^{\mu(R_1)}R_1 M\kappa^{-1/2}\int_{R_0}^{R_1}\frac{d\eta}{\eta^2}&\nonumber\\
&\displaystyle 
\leq (\Omega+1)e^{\mu(R_1)}M\kappa^{-1/2}(\frac{R_1}{R_0}-1)&\nonumber\\
&\displaystyle \leq (\Omega+1)M\sqrt{1-2M/R_1}\kappa^{-1/2}\epsilon.\label{S2uniform}& 
\end{eqnarray}
To obtain a similar form of the inequality (\ref{original}) also when $S_2\neq 0$ we simply use that 
\begin{equation}
\frac{2}{1+\sqrt{1-2M/R_1}}\geq 1,\label{sameform}
\end{equation}
and with (\ref{S2uniform}) we thus get
\begin{equation}
M\leq\frac{2(\Omega+1)M(1+\kappa^{-1/2}\epsilon)\sqrt{1-2M/R_1}}{1+\sqrt{1-2M/R_1}}.\label{ineqM}
\end{equation} 
This yields
\[ 
\frac{2M}{R_1}\leq\frac{(2\Omega+1+2(\Omega+1)\kappa^{-1/2}\epsilon)^{2}-1}{(2\Omega+1+2(\Omega+1)\kappa^{-1/2}\epsilon)^{2}}.
\]
Hence, a uniform bound on $1/(1-2m/r)$ implies that the limit supremum of $2M^j/R_1^j$, as $j\to\infty$, is bounded by $$\frac{(2\Omega+1)^2-1}{(2\Omega+1)^2}.$$ A uniform bound is given by Theorem 2, which we prove below, and which completes the proof of Theorem 1. 
\begin{flushright}
$\Box$
\end{flushright}
\textit{Proof of Theorem 1: }
Let us first show that the second statement in the theorem is a consequence of the first. Since $m(r)=0$ when $r<R_0$ and $m(r)=M$ when $r>R_1$ it is sufficient to take $r\in [R_0,R_1].$ If $\epsilon\leq\kappa/2,$ we then have 
\begin{eqnarray*}
&2m/r\leq 2M/r\leq 2M/R_0\leq R_1(1-\kappa)/R_0&\\
&\leq (1+\epsilon)(1-\kappa)<1-\kappa+\epsilon\leq 1-\kappa /2.&
\end{eqnarray*}
Thus we only have to prove that $2M/R_1\leq 1-\kappa.$ 
The proof is by contradiction. Hence, assume that $2M/R_1> 1-\kappa,$ and let 
\[ 
\eta^{*}:=\min\{\eta\in [R_0,R_1]:\frac{2m(\eta)}{\eta} = 1-\kappa\}. 
\] 
Consider (\ref{Minequality}) again. The term $S_1$ above can be estimated as before (note that the result in \cite{BR} shows that $2m/r< 1$ for a given solution so there are no regularity problems) but for $S_2$ we split the integration over $[R_0,\eta^{*}]$ and $[\eta^{*},R_1].$ The integral over the former interval is estimated as in (\ref{S2uniform}) and we get 
\[
(\Omega+1)e^{\mu(R_1)}R_1\int_{R_0}^{\eta^{*}}\frac{m(\eta)}{\eta^2 \sqrt{1-2m(\eta)/\eta}} d\eta\leq ... \leq (\Omega+1)M\sqrt{1-2M/R_1}\frac{\epsilon}{\sqrt{\kappa}}. 
\]
If it now holds that 
\begin{equation}
\int_{\eta^{*}}^{R_1}\frac{m(\eta)}{\eta^2 \sqrt{1-2m(\eta)/\eta}}\,d\eta\leq \frac{KM}{R_1}, \label{assumptionK}
\end{equation}
then the estimate for $S_2$ reads 
\[
S_2\leq (\Omega+1)M\sqrt{1-2M/R_1}\frac{\epsilon}{\sqrt{\kappa}}+(\Omega+1)KM\sqrt{1-2M/R_1}. 
\]
This leads to an inequality similar to (\ref{ineqM}). More precisely we obtain 
\begin{eqnarray}
\displaystyle\frac{2M}{R_1}&\leq&\frac{(2\Omega+1+\frac{2(\Omega+1)^2}{2\Omega+1}(K+\epsilon/\sqrt{\kappa}))^2-1}{(2\Omega+1+\frac{2(\Omega+1)^2}{2\Omega+1}(K+\epsilon/\sqrt{\kappa}))^{2}}\nonumber\\
\displaystyle &\leq& 1-\frac{1}{(2\Omega+1+\frac{2(\Omega+1)^2}{2\Omega+1}(K+a))^{2}}=1-\kappa,\label{AO} 
\end{eqnarray}
where we used that $\epsilon\leq a\kappa^{1/2},$ together with a slightly different form of (\ref{sameform}) using $2M/R_1\geq 1-\kappa>8/9$, i.e. 
\begin{equation}
\frac{4}{3(1+\sqrt{1-2M/R_1})}\geq 1.
\end{equation}
Now we have assumed that (\ref{AO}) is not true, so in fact the following inequality must hold 
\begin{equation}
\int_{\eta^{*}}^{R_1}\frac{m(\eta)}{\eta^2 \sqrt{1-2m(\eta)/\eta}}\,d\eta>\frac{KM}{R_1}.\label{assumptionKc} 
\end{equation}
Let us now consider (\ref{Minequality}) again. We will show that (\ref{assumptionKc}), together with the hypotheses in the theorem, imply that the right hand of this inequality is less than $M,$ which gives a contradiction. 
First we need a couple of facts. Since 
\[
\int_{r}^{R_1}(4\pi\eta\rho-\frac{m}{\eta^2})e^{\lambda}\, d\eta=
-\int_{r}^{R_1}\frac{d}{d\eta}e^{-\lambda}\, d\eta=e^{-\lambda(r)}-e^{-\lambda(R_1)}, 
\]
we get 
\begin{equation}
\int_{r}^{R_1}4\pi\eta\rho e^{\lambda}\, d\eta=e^{-\lambda(r)}-e^{-\lambda(R_1)}+\int_{r}^{R_1}\frac{m}{\eta^2}e^{\lambda}\, d\eta. \label{duality}
\end{equation}
Since $\lambda\geq 0$ and $\rho=p=0$ when $r>R_1,$ we have 
\begin{equation}
(\mu+\lambda)(r)=-\int_{r}^{R_1}4\pi\eta(\rho+p)e^{2\lambda}\, d\eta\leq
-\int_{r}^{R_1}4\pi\eta\rho e^{\lambda}\, d\eta. \label{expest}
\end{equation}
Now we split the integral in the inequality for $M$ in two parts 
\begin{eqnarray}
(\Omega+1)\int_{R_0}^{R_1}4\pi\eta^2 e^{\mu+\lambda}\rho\, d\eta&=&(\Omega+1)\int_{R_0}^{\eta^{*}}4\pi\eta^2 e^{\mu+\lambda}\rho\,d\eta\nonumber\\
&+&(\Omega+1)\int_{\eta^{*}}^{R_1}4\pi\eta^2 e^{\mu+\lambda}\rho\, d\eta\nonumber\\
&=&:T_1+T_2. 
\end{eqnarray}
Using the two facts above we get 
\begin{eqnarray}
\displaystyle T_1&=&(\Omega+1)\int_{R_0}^{\eta^{*}}4\pi\eta^2 e^{\mu+\lambda}\rho\,d\eta\leq 
(\Omega+1)\int_{R_0}^{\eta^{*}}4\pi\eta^2\rho e^{-\int_{\eta}^{R_1}4\pi\sigma\rho e^{\lambda}\,d\sigma}
d\eta\nonumber\\
\displaystyle &=&(\Omega+1)\int_{R_0}^{\eta^{*}}4\pi\eta^2\rho e^{e^{-\lambda(R_1)}-e^{-\lambda(\eta)}-
\int_{\eta}^{R_1}m(\sigma)e^{\lambda}/\sigma^{2}\,d\sigma}d\eta. 
\end{eqnarray}
We have $e^{-\lambda(R_1)}\leq e^{-\lambda(\eta)},$ since $\eta\leq\eta^{*},$ and $2M/R_1>1-\kappa.$ We thus get due to (\ref{assumptionKc}), again since $\eta\leq\eta^{*},$ 
\begin{equation}
T_1\leq (\Omega+1)\int_{R_0}^{\eta^{*}}4\pi\eta^2\rho e^{-KM/R_1}d\eta\leq (\Omega+1)M e^{-KM/R_1}.\label{T1}
\end{equation}
Note that $M/R_1> 4/9$ so that $T_1$ is small if $K$ is large. 
Let $\sigma^{*}\in [\eta^{*},R_1]$ be such that 
\[
e^{2\lambda(\sigma^{*})}\leq e^{2\lambda(\sigma)},\;\forall\sigma\in [\eta^{*},R_1].
\]
Using again that 
\[
(\mu+\lambda)(\eta)=-\int_{\eta}^{R_1}4\pi\sigma(\rho+p)e^{2\lambda}\, d\sigma, 
\]
we get 
\begin{eqnarray}
\displaystyle T_2&\leq&(\Omega+1)R_1\int_{\eta^{*}}^{R_1}4\pi\eta e^{\mu+\lambda}\rho\, d\eta\nonumber\\
\displaystyle&\leq&(\Omega+1)R_1
\int_{\eta^{*}}^{R_1}4\pi\eta\rho e^{-e^{2\lambda(\sigma^{*})}\int_{\eta}^{R_1}4\pi\sigma(\rho+p)\, d\sigma}\, d\eta\nonumber\\
\displaystyle &\leq& (\Omega+1)R_1
\int_{\eta^{*}}^{R_1}4\pi\eta\rho e^{-e^{2\lambda(\sigma^{*})}\int_{\eta}^{R_1}4\pi\sigma\rho\, d\sigma}\, d\eta\nonumber\\
&=&(\Omega+1)R_1e^{-2\lambda(\sigma^{*})}\int_{\eta^{*}}^{R_1}\frac{d}{d\eta}\left(e^{-e^{2\lambda(\sigma^{*})}\int_{\eta}^{R_1}4\pi\sigma\rho\, d\sigma}\right)\, d\eta\nonumber\\
\displaystyle &=& (\Omega+1)R_1e^{-2\lambda(\sigma^{*})}\left(1-e^{-e^{2\lambda(\sigma^{*})}\int_{\eta^{*}}^{R_1}4\pi\sigma\rho\, d\sigma}\right)\nonumber\\
&\leq&(\Omega+1)R_1e^{-2\lambda(\sigma^{*})}=(\Omega+1)R_1(1-2m(\sigma^{*})/\sigma^{*}).\label{T2p}
\end{eqnarray}
Now since $m(r)$ is increasing we have for any $\sigma\in [\eta^{*},R_1],$ 
\begin{eqnarray*}
&\displaystyle\frac{2m(\sigma)}{\sigma}\geq \frac{2m(\eta^{*})}{\sigma}\geq \frac{2m(\eta^{*})}{R_1}= \frac{\eta^{*}}{R_1}\frac{2m(\eta^{*})}{\eta^{*}}&\\
&\displaystyle\geq\frac{R_0}{R_1}\frac{2m(\eta^{*})}{\eta^{*}}= \frac{R_0}{R_1}(1-\kappa)\geq\frac{1-\kappa}{1+\epsilon}.&
\end{eqnarray*}
Using this estimate twice we obtain 
\begin{eqnarray*}
&\displaystyle T_2\leq (\Omega+1)R_1(1-2m(\sigma^{*})/\sigma^{*})\leq (\Omega+1)R_1(1-\frac{1-\kappa}{1+\epsilon})\\
&\displaystyle =(\Omega+1)\frac{R_1}{M}M\frac{\kappa+\epsilon}{1+\epsilon}\leq 2(\Omega+1)M\frac{(1+\epsilon)}{(1-\kappa)}\frac{(\kappa+\epsilon)}{(1+\epsilon)}\\
&\displaystyle =\frac{2(\Omega+1)M(\kappa+\epsilon)}{1-\kappa}.&
\end{eqnarray*}
In conclusion the following inequality holds, using again that $M/R_1> 4/9,$ together with the condition on $\epsilon,$ 
\begin{eqnarray*}
M&\leq&(\Omega+1)M e^{-KM/R_1}+\frac{2(\Omega+1)M(\kappa+\epsilon)}{1-\kappa}\\
&\leq& (\Omega+1)M e^{-4K/9}+\frac{2(\Omega+1)M(\kappa+\epsilon)}{1-\kappa}<M.
\end{eqnarray*}
Hence we have obtained a contradiction and the proof is complete. 
\begin{flushright}
$\Box$
\end{flushright}  
\begin{center}
\textbf{Acknowledgement}
\end{center}
The author wants to thank Gerhard Rein for reading and commenting the manuscript, and Alan Rendall for carefully answering my questions. 

\end{document}